\documentclass[sigconf,natbib=true,nonacm]{acmart}

\setcopyright{none}
\renewcommand\footnotetextcopyrightpermission[1]{}
\acmYear{2026}
\acmISBN{}
\acmDOI{}

\usepackage{booktabs}
\usepackage{amsmath}
\usepackage{graphicx}
\usepackage{tikz}
\usetikzlibrary{positioning, arrows.meta, shapes.geometric,
                fit, backgrounds, calc}

\begin{document}

\title{A Triple-Robustness Analysis of Retrieval-Augmented Generation
  for Multi-Hop Requirements Traceability}

\author{Meftun Akarsu}
\affiliation{%
  \institution{Turkish Aerospace Industries}
  \country{}
}

\author{Burak \"Ozdemir}
\affiliation{%
  \institution{Turkish Aerospace Industries}
  \country{}
}

\author{Do\u{g}ancan B\"uy\"uk\c{c}olak}
\affiliation{%
  \institution{Turkish Aerospace Industries}
  \country{}
}

\author{Recep Kaan Karaman}
\authornote{Corresponding author: kaan.karaman@tum.de}
\affiliation{%
  \institution{Technical University of Munich}
  \country{}
}

\begin{abstract}
Reported verdicts on GraphRAG versus vector RAG disagree, and the
evidence is typically tied to a single corpus, embedder, and judge --
and, we show, to where citation quality is measured. We present a
triple-robustness analysis that holds a five-pipeline architecture
matrix fixed and varies embedder (local e5-small vs.\ Azure
text-embedding-3-small), corpus (DO-178C typed-edge requirements vs.\
Wikipedia paragraph chains via MuSiQue), and judge (paired GPT-5.4
$\times$ GPT-4.1 on both corpora), over $2{\times}4{,}440$ main-matrix
runs, 600 cross-corpus runs, and over 5{,}000 faithfulness judgments.
\textbf{(C2a)} GraphRAG's graph walk floods the context window at
precision 0.12--0.23, but the synthesizer cites selectively at
precision 0.48--0.65; scoring the retrieved set as the attribution set
inverts the architecture ranking, which reconciles part of the
disagreement in prior reports. \textbf{(C1)} Answer-level citation
winners are corpus- and stratum-conditional but embedder-robust:
GraphRAG ties vanilla on short-hop DO-178C queries and wins every
MuSiQue stratum, while agentic pipelines lead only on 3+-hop
requirements queries. \textbf{(C2b)} Faithfulness is
corpus-conditional: on DO-178C it declines with hop distance (trend
$p{<}0.05$ in three of four judge$\times$embedder combinations); on
Wikipedia chains neither judge shows a collapse. \textbf{(C3)}
Single-judge LLM faithfulness is fragile to retrieval state: GPT-5.4's
self-$\kappa$ across embedders is $0.137$ (41\% verdict change)
against a same-day test--retest floor of $0.76$, and re-judging frozen
inputs eleven weeks later gives $\kappa{\leq}0.14$ for both judges. A learned router on
dense embeddings alone reaches macro-$F_1$ $0.86$ on hop
classification \textbf{(C4)}. We argue that RAG architecture claims
should be tested at this level of robustness -- including robustness
to the citation-measurement point -- before they are trusted.
\end{abstract}



\maketitle

\section{Introduction}
\label{sec:intro}

DO-178C-style aerospace requirements are authored as typed link graphs:
each requirement \textit{derives\_from} parents, \textit{satisfies}
system-level intent, and \textit{traces\_to} verification artifacts.
Certification authorities increasingly require auditable chains spanning
two or more hops~\cite{do178c,arp4754a,easa2025npa,faa2024airoadmap}.
RAG architectures for this task have produced contradictory verdicts:
Microsoft GraphRAG~\cite{graphrag2024} exploits structure but
under-performs vector RAG on detailed queries~\cite{ragvsgraphrag2025,
graphragbench2026}; Adaptive-RAG~\cite{adaptiverag2024} routes by
complexity yet is hop-blind; LiSSA~\cite{lissa2025} and
TVR~\cite{niu_tvr2025} omit typed-graph reasoning. Prior evaluations
report which architecture wins in a given setting, but not why the
outcome changes when the setting changes.

We address this with a \textbf{triple-robustness} analysis. Holding a
five-pipeline architecture matrix fixed (vanilla, agentic, agentic+graph,
GraphRAG, learned adaptive), we vary three orthogonal axes: the retrieval
embedder (e5-small 384d $\to$ Azure 3-small 1536d), the corpus (DO-178C
$\to$ Wikipedia paragraph chains via MuSiQue), and the faithfulness
judge (paired GPT-5.4 $+$ GPT-4.1 on both corpora). The design produces $2{\times}4{,}440$ main-matrix runs (one
full matrix per embedder), 600 cross-corpus runs plus a 200-run
distractor-edge control, and over 5{,}000 faithfulness judgments.

We make four contributions.
\textbf{(C2a)} GraphRAG's walk and its synthesizer pull in opposite
directions: the walk fills the context at precision 0.12--0.23, the
answer cites 3--5 IDs at precision 0.48--0.65
(Table~\ref{tab:pathology}). Scoring the retrieved set as the
attribution set -- a choice some comparisons make implicitly -- ranks
GraphRAG last; scoring the answer's citations ranks it first or
tied-first.
\textbf{(C2b)} The faithfulness consequence of context flooding is
\emph{corpus-conditional}: DO-178C declines across hops
(74\%$\to$40\% local); MuSiQue shows no collapse under either judge.
\textbf{(C1)} Answer-level winners are
\emph{corpus- and stratum-conditional but embedder-robust}
(Table~\ref{tab:dominance}): vanilla and GraphRAG tie on 1--2-hop
DO-178C queries, graph-aided pipelines lead on 3+-hop, and GraphRAG
wins every MuSiQue stratum.
\textbf{(C3)} \emph{Single-judge faithfulness is unstable}: GPT-5.4
self-$\kappa$ across embedders is $0.137$ against a $0.76$
test--retest floor (Table~\ref{tab:judges}).
\textbf{(C4, supporting)} a logistic-regression router on dense embeddings
alone reaches macro-$F_1$ $0.86$ on hop classification.

\section{Related Work}
\label{sec:related}

\noindent\textbf{Agentic and Adaptive RAG.}
Self-RAG~\cite{selfrag2024} and CRAG~\cite{crag2024} introduced reflective
and corrective retrieval. Adaptive-RAG~\cite{adaptiverag2024} routes by
predicted query complexity; Probing-RAG~\cite{probingrag2025} extends with
internal-state probes; Search-R1~\cite{searchr1_2025} trains agentic
retrieval via RL; RAG-Critic~\cite{ragcritic2025} adds an iterative critic.
None condition routing on typed-graph hop distance or expose a typed
graph-lookup tool.

\noindent\textbf{GraphRAG and multi-hop QA.}
Microsoft GraphRAG~\cite{graphrag2024} and follow-ups~\cite{lightrag2024,
hipporag2_2025,graphr1_2025} construct entity-relation graphs at indexing
time. Han et~al.~\cite{ragvsgraphrag2025} and GraphRAG-Bench~\cite{graphragbench2026}
report that GraphRAG does not dominate vector RAG across all query
types. Our C2a shows one reason such verdicts disagree: whether
citation quality is scored over the retrieved set or over the answer's
citations decides which side wins, and C2b separates the
context-flooding mechanism from its faithfulness consequence across
corpora. MuSiQue~\cite{musique2022} and
MultiHop-RAG~\cite{multihoprag2024} are adjacent multi-hop benchmarks; we
use MuSiQue for cross-corpus replication of C2a and C2b.

\noindent\textbf{RAG for requirements traceability.}
LiSSA~\cite{lissa2025}, TVR~\cite{niu_tvr2025}, and Graph-RAG for
compliance~\cite{graphragcompliance2024} evaluate single regulated domains
without hop stratification or dual-judge protocols.

\noindent\textbf{Citation evaluation, judge bias, and the $\kappa$ paradox.}
ALCE~\cite{alce2023} defined citation $P/R/F_1$ over generated
answers; Wallat~et~al.~\cite{wallat2024} show that ALCE-style
correctness is not faithfulness, motivating LLM judges. RAGChecker~\cite{ragchecker2024}
provides a single-judge framework. Self-preference
bias~\cite{selfpref_bias2024} motivates judge ensembles, and the
$\kappa$ paradox of Feinstein and Cicchetti~\cite{feinstein1990},
with Gwet's AC1~\cite{gwet2008}, establishes the prevalence
correction we apply. Our C3 adds a
statistic this literature does not track: the judge's agreement with
itself under retrieval-state and date changes.

\section{Method}
\label{sec:method}

\begin{figure}[t!]
  \centering
  \resizebox{0.74\columnwidth}{!}{

\begin{tikzpicture}[
  node distance=4mm,
  >={Stealth[length=2mm]},
  every node/.style={font=\sffamily\scriptsize}
]
\tikzset{
  box/.style    ={draw=black, line width=0.5pt, rounded corners=2pt,
                  align=center, inner sep=2pt},
  light/.style  ={box, fill=black!8},
  mid/.style    ={box, fill=black!18},
  pipe/.style   ={light, text width=1.55cm, minimum width=1.7cm,
                  minimum height=7.5mm},
  lp/.style     ={light, text width=1.45cm, minimum width=1.55cm,
                  minimum height=6.5mm},
  store/.style  ={cylinder, shape border rotate=90, aspect=0.25,
                  draw=black, line width=0.5pt, fill=black!18,
                  align=center, inner sep=2pt,
                  text width=1.7cm, minimum height=7mm},
  arr/.style    ={->, line width=0.4pt},
  darr/.style   ={arr, dashed},
  lab/.style    ={font=\sffamily\tiny, fill=white, inner sep=1pt},
  group/.style  ={draw=black, line width=0.5pt, rounded corners=2pt,
                  dotted, inner sep=4pt}
}

\node[light, text width=1.7cm] (q) at (0,0) {Query $q$};
\node[mid, below=4mm of q, text width=4.4cm] (router)
  {Hop-Adaptive Router\\{\tiny predicts $\widehat{H}(q) \in \{1, 2, 3{+}\}$}};
\draw[arr] (q) -- (router);

\node[pipe, below=5mm of router, xshift=-2.775cm] (p1)
  {Vanilla\\{\tiny embed $\to$ top-$k$ $\to$ LLM}};
\node[pipe, right=1.5mm of p1] (p2)
  {Agentic\\{\tiny LangGraph loop}};
\node[pipe, right=1.5mm of p2] (p3)
  {GraphRAG\\{\tiny seeds + 2-hop walk}};
\node[pipe, right=1.5mm of p3] (p4)
  {Agentic+Graph\\{\tiny LangGraph + graph}};
\foreach \i in {1,2,3,4}
  \draw[arr] (router.south) -- (p\i.north);

\node[lp] (critic) at (-1.7,-4.6) {Router/Critic};
\node[lp, right=1.5mm of critic] (retr)  {Retriever};
\node[lp, right=1.5mm of retr]   (tools) {Tools\\{\tiny search, graph}};
\node[lp, below=4mm of retr]     (synth) {Synthesizer};

\draw[arr] (critic) -- (retr);
\draw[arr] (retr)   -- (tools);
\draw[arr] (tools.north) to[out=90,in=90,looseness=0.5]
  node[lab, midway, yshift=1mm] {iter $\le 3$} (critic.north);
\draw[arr] (critic.south) |- (synth.west);

\begin{scope}[on background layer]
  \node[group, fit=(critic)(retr)(tools)(synth)] (loop) {};
\end{scope}
\node[lab, anchor=north west]
  at ([xshift=2pt,yshift=-1pt]loop.north west) {Agentic Loop};

\draw[darr] (p2.south) -- ++(0,-2mm) -| (loop.north west)
  node[lab, pos=0.4, above] {uses};
\draw[darr] (p4.south) -- ++(0,-2mm) -| (loop.north east)
  node[lab, pos=0.4, above] {uses};

\node[light, below=6mm of synth, text width=4.5cm] (out)
  {Answer $a(q)$ + cited IDs $C(q)$};
\draw[arr] (synth.south) -- (out.north);

\draw[arr] (p1.south) -- ++(0,-2mm) -|
  ([xshift=-4mm]loop.south west) |- (out.west);
\draw[arr] (p3.south) -- ++(0,-2mm) -|
  ([xshift=4mm]loop.south east)  |- (out.east);

\node[store, below=10mm of out, xshift=-22mm] (s1)
  {Vector Store\\{\tiny (Chroma)}};
\node[store, below=10mm of out, xshift=22mm]  (s2)
  {Graph Store\\{\tiny (Neo4j Aura)}};

\draw[darr] (s1.north) -- ++(0,4mm)
  node[lab, anchor=south] {$\to$ P1--P4};
\draw[darr] (s2.north) -- ++(0,4mm)
  node[lab, anchor=south] {$\to$ P3, P4};

\end{tikzpicture}}%
  \caption{Five-pipeline architecture with shared embedder, vector store,
    and typed-edge graph. Vanilla and GraphRAG bypass the agentic loop;
    Agentic and Agentic+Graph route through router/critic--retriever--tools
    (iter${\leq}3$) before synthesis. The adaptive pipeline (V1 rule-based
    or V2 learned) selects one of the four base pipelines per query.}
  \Description{Block diagram of the five-pipeline RAG architecture: a
    query enters a hop-adaptive router that selects one of four base
    pipelines (Vanilla, Agentic, GraphRAG, Agentic+Graph). The Agentic
    and Agentic+Graph pipelines invoke an inner Agentic Loop comprising
    a router/critic, retriever, tools (search and graph), and a
    synthesizer that iterates up to three times. Outputs combine into
    an answer with cited IDs. A shared vector store (Chroma) feeds
    P1--P4 and a shared graph store (Neo4j Aura) feeds P3 and P4.}
  \label{fig:architecture}
\end{figure}
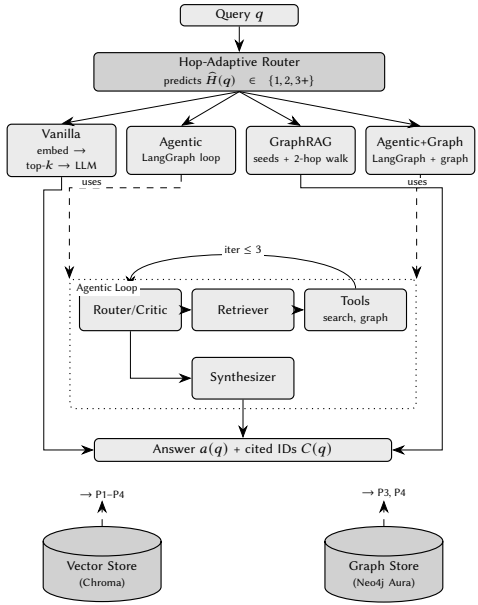

\subsection{Pipelines}
\label{sec:pipelines}
All pipelines share embedder, ChromaDB vector store, Neo4j typed-edge
graph, Azure GPT-5.4 generator, and grounded synthesis prompt; they
differ only in retrieval. \textbf{(i)~Vanilla}: dense top-10 retrieval,
top-5 context, one synthesis call. \textbf{(ii)~Agentic}: LangGraph
router-retriever-critic loop, \texttt{search\_documents} tool only, iter
cap 3. \textbf{(iii)~GraphRAG}: vector seeds (8) + up-to-2-hop typed-graph
walk via Cypher (walk cap 30, context cap 15), single synthesis. This
is a typed-edge local-walk retriever in the GraphRAG family, not Edge
et~al.'s community-summarization system~\cite{graphrag2024}; we use
``GraphRAG'' as the family label. \textbf{(iv)~Agentic+Graph}: (ii) plus
a typed-edge \texttt{graph\_lookup} tool. \textbf{(v)~Adaptive}: chooses
among (i)--(iv) per query; we evaluate a rule-based router (V1) and a
learned out-of-fold logistic regression (V2).

\subsection{Triple-Robustness Axes}
\label{sec:axes}
\textbf{Embedder:} local intfloat/multilingual-e5-small (384d) vs.\ Azure
\texttt{text-embedding-3-small} (1536d); ChromaDB collections rebuilt per
embedder, Neo4j shared. \textbf{Corpus:} a synthetic 1{,}132-requirement
DO-178C-style aerospace certification corpus across 32 modules,
generated by GPT-5.4 from a hand-authored module taxonomy and
publicly released; and a 200-query
MuSiQue~\cite{musique2022} subset (67/67/66 across 2/3/4-hop, mapped to
our 1/2/3+-hop strata, \textit{REFERENCES} edges only between consecutive
supporting paragraphs). \textbf{Judge:} GPT-5.4 $+$ GPT-4.1 on both the
DO-178C main matrix and MuSiQue.

\subsection{Metrics and Judging}
\label{sec:metrics}
We report ALCE-style citation $P/R/F_1$~\cite{alce2023} against
gold ID sets. Cited IDs are parsed from the answer text with a
vocabulary-anchored parser (exact corpus IDs, longest match first);
hallucinated IDs that reuse a real module prefix count against
precision, and incidental tokens (SHA-256, DO-254) are excluded. We
separately report \emph{context precision} -- the gold fraction of the
retrieved set handed to the synthesizer -- and retrieval recall, since
conflating the context set with the answer's citations changes which
architecture appears to win (\S\ref{sec:c2}). Each
faithfulness judgment is a strict-JSON binary verdict over the
retrieved-context block. We report Cohen's $\kappa$, Gwet's
AC1~\cite{gwet2008,feinstein1990}, raw agreement, and McNemar
exact-binomial $p$; per faithfulness cell, Wilson 95\% CIs and
Cochran--Armitage hop-trend tests; and three same-judge controls
(test--retest, embedder swap, eleven-week re-judge) on a paired
300-tuple subset (\S\ref{sec:c3}).

\subsection{V2 Learned Router}
\label{sec:routerdesign}
14 features per query: 11 hand text features (ID regex, keyword flags,
log token length) and 3 PCs of the query embedding (fit per fold).
Multinomial logistic regression, $L_2$ ($C{=}1$), Platt-calibrated,
under stratified 10-fold $\times$ 5-repeat CV (50 fits). The per-stratum
routing target is the empirical mean-$F_1$ winner, derived once from
the full locked matrix (not re-fit per fold); only hop prediction is
out-of-fold, reported for all 296 queries.

\section{Experimental Setup}
\label{sec:experiments}

\textbf{Main matrix and cross-corpus sample.} DO-178C: 5 pipelines $\times$
296 hop-stratified queries $\times$ 3 seeds $= 4{,}440$ runs, executed
under both embedders (\emph{v2}, \emph{v3}). MuSiQue: 3 pipelines (vanilla,
GraphRAG, agentic+graph) $\times$ 200 queries $\times$ 1 seed $= 600$ runs,
Azure embedder only, plus a 200-run GraphRAG distractor-edge control
(\S\ref{sec:c1}) on the same queries and vector seeds. Agentic and
adaptive are not re-run on MuSiQue (C2a needs only the three-pipeline
trio).

\textbf{Faithfulness judging.} On DO-178C, the multi-judge protocol
(GPT-5.4 $+$ GPT-4.1) is applied to two disjoint 300-row stratified
batches (60 per pipeline; seeds 42 and 43), the second judged eleven
weeks after the first, yielding 1{,}200 paired binary judgments per
embedder. Each batch is \emph{pinned} to the same 300 (query,
pipeline, repeat) tuples across embedders so C3 deltas are computed
on identical items; batches are reported separately, never pooled
across judging dates. On MuSiQue, we dual-judge all 600 rows.
Same-input controls re-judge 300 v2 tuples twice with GPT-5.4
(test--retest) and once eleven weeks later (temporal drift); a
generator-swap control re-synthesizes 332 v3 rows with GPT-4.1 on
frozen retrievals and dual-judges them.

\textbf{Statistical protocol.} 95\% BCa bootstrap
intervals~\cite{du2025bootstrap} ($B{=}1000$, paired at the query
level), Wilcoxon signed-rank with Holm correction across the 30
pipeline-pair $\times$ stratum
contrasts~\cite{bergkirkpatrick2012,koehn2004}, and Cliff's $\delta$
(negligible ${<}0.147$). A pipeline pair
is reported as significantly different only when the BCa interval excludes
zero, Holm $p{<}0.05$, and $|\delta|\!\geq\!0.147$ jointly hold.

\textbf{Reproducibility.} The full pipeline regenerates from a
single make target on locked CSVs. Azure model snapshots:
\texttt{gpt-5.4-}\allowbreak\texttt{2026-03-05} (generator + judge),
\texttt{gpt-4.1-2025-04-14} (judge); reasoning-mode decoding with an
8{,}192-token completion cap. The MuSiQue
subgraph (3{,}996 paragraph chunks, 399 \textit{REFERENCES} edges) is
built deterministically from the \texttt{dgslibisey}
MuSiQue mirror (validation split, seed $20260511$).

\section{Results}
\label{sec:results}

\begin{table*}[t]
  \centering
  \caption{Per-stratum citation $F_1$ across three settings. v2 main = DO-178C synthetic with local e5-small embeddings (4{,}440 runs); v3 main = same corpus with Azure text-embedding-3-small (4{,}440 runs); MuSiQue = 200-query Wikipedia stratified subset with Azure embedder (600 runs, vanilla / agentic-graph / graphrag only). Bold marks the per-(setting, stratum) maximum.}
  \label{tab:dominance}
  \small
  \begin{tabular}{l ccc | ccc | ccc}
    \toprule
     & \multicolumn{3}{c|}{v2 main (local)} & \multicolumn{3}{c|}{v3 main (Azure)} & \multicolumn{3}{c}{MuSiQue (Azure)} \\
    System & 1h & 2h & 3+h & 1h & 2h & 3+h & 1h & 2h & 3+h \\
    \midrule
    vanilla & \textbf{0.757} & 0.712 & 0.089 & \textbf{0.756} & 0.689 & 0.195 & 0.742 & 0.415 & 0.277 \\
    agentic & 0.541 & 0.414 & 0.175 & 0.533 & 0.430 & 0.203 & --- & --- & --- \\
    agentic-graph & 0.566 & 0.415 & \textbf{0.219} & 0.547 & 0.421 & 0.251 & 0.457 & 0.296 & 0.249 \\
    graphrag & 0.751 & \textbf{0.720} & 0.172 & 0.745 & \textbf{0.707} & 0.253 & \textbf{0.841} & \textbf{0.632} & \textbf{0.463} \\
    adaptive & 0.630 & 0.499 & 0.181 & 0.651 & 0.492 & \textbf{0.254} & --- & --- & --- \\
    \bottomrule
  \end{tabular}
\end{table*}

\begin{figure}[t!]
  \centering
  \includegraphics[width=0.88\columnwidth]{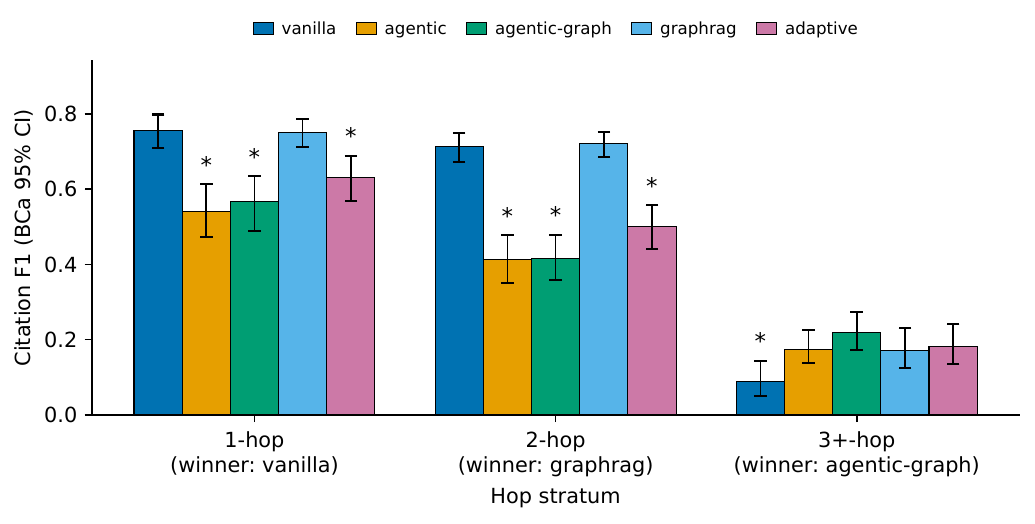}
  \caption{Per-stratum citation $F_1$ on the v2 main matrix (DO-178C,
    local e5-small embedder): five pipelines with BCa 95\% CIs;
    \mbox{$*$} marks pipelines significantly different from the
    per-stratum winner (Holm-Wilcoxon $p{<}0.05$,
    $|\delta|{\geq}0.147$).}
  \Description{Grouped bar chart of per-stratum citation F1 for five
    RAG pipelines (vanilla, agentic, agentic-graph, graphrag, adaptive)
    on the DO-178C v2 main matrix, broken down by 1-hop, 2-hop, and
    3+-hop strata, with error bars and asterisks marking pipelines
    significantly different from the per-stratum winner.}
  \label{fig:perstratum}
\end{figure}

\subsection{C1: Winners are Corpus-Conditional, Embedder-Robust}
\label{sec:c1}
Table~\ref{tab:dominance} reports per-stratum $F_1$ across three settings.
On DO-178C the pattern repeats under both embedders: vanilla and
GraphRAG are statistically tied on 1-hop and 2-hop (Wilcoxon
$p{=}0.12/0.65$ local, negligible $\delta$), the agentic loop costs
0.19--0.30 $F_1$ on those strata, and on 3+-hop the ordering inverts --
vanilla drops to last and the graph-aided pipelines lead (local
embedder: agentic-graph 0.219 vs.\ vanilla 0.089, Holm
$p{<}10^{-6}$; under the Azure embedder the same ordering holds but no
3+-hop pair passes the joint significance criterion). On MuSiQue
(three-pipeline field) GraphRAG wins every stratum outright
($F_1{=}0.841/0.632/0.463$; vs.\ vanilla, Wilcoxon
$p{\leq}5{\times}10^{-4}$, $\delta{=}0.27$--$0.43$). Because the
MuSiQue graph's \textit{REFERENCES} edges connect only gold supporting
paragraphs, we re-ran the GraphRAG arm on a control graph with
consecutive-distractor edges added (context precision falls
$0.23{\to}0.14$; the walk cap is always reached). GraphRAG loses only
$0.02/0.05/0.09$ $F_1$ per stratum and still beats vanilla everywhere
($+0.08/+0.16/+0.10$, $p{\leq}0.014$): the MuSiQue win is not an
artifact of gold-only edges. Which architecture wins is thus corpus-
and stratum-conditional, but stable under embedder swap.

\begin{table}[t]
  \centering
  \caption{GraphRAG context flooding vs.\ citation behavior (C2a) and faithfulness by stratum (C2b), across embedder and corpus swaps. The graph walk fills the context window with low-precision material, but the synthesizer cites only a third of it at much higher precision. C2b cells are GPT-5.4\,/\,GPT-4.1 faithful rates.}
  \label{tab:pathology}
  \small
  \begin{tabular}{l ccc}
    \toprule
    \textit{C2a: context vs.\ citation} & v2 main & v3 main & MuSiQue \\
    \midrule
    Mean context IDs & 14.9 & 14.9 & 11.2 \\
    Context precision & 0.120 & 0.129 & 0.227 \\
    Mean IDs cited in answer & 4.9 & 5.0 & 3.4 \\
    Citation precision & 0.480 & 0.493 & 0.654 \\
    Retrieval recall & 0.681 & 0.724 & 0.873 \\
    Citation $F_1$ overall & 0.551 & 0.571 & 0.646 \\
    \midrule
    \textit{C2b: faithfulness} & 1-hop & 2-hop & 3+-hop \\
    \midrule
    v2 main & 0.74\,/\,0.78 & 0.64\,/\,0.82 & 0.40\,/\,0.33 \\
    v3 main pinned & 0.52\,/\,0.96 & 0.55\,/\,0.86 & 0.40\,/\,0.20 \\
    MuSiQue & 0.42\,/\,0.94 & 0.54\,/\,0.88 & 0.58\,/\,0.80 \\
    \bottomrule
  \end{tabular}
\end{table}

\begin{figure}[t!]
  \centering
  \includegraphics[width=0.84\columnwidth]{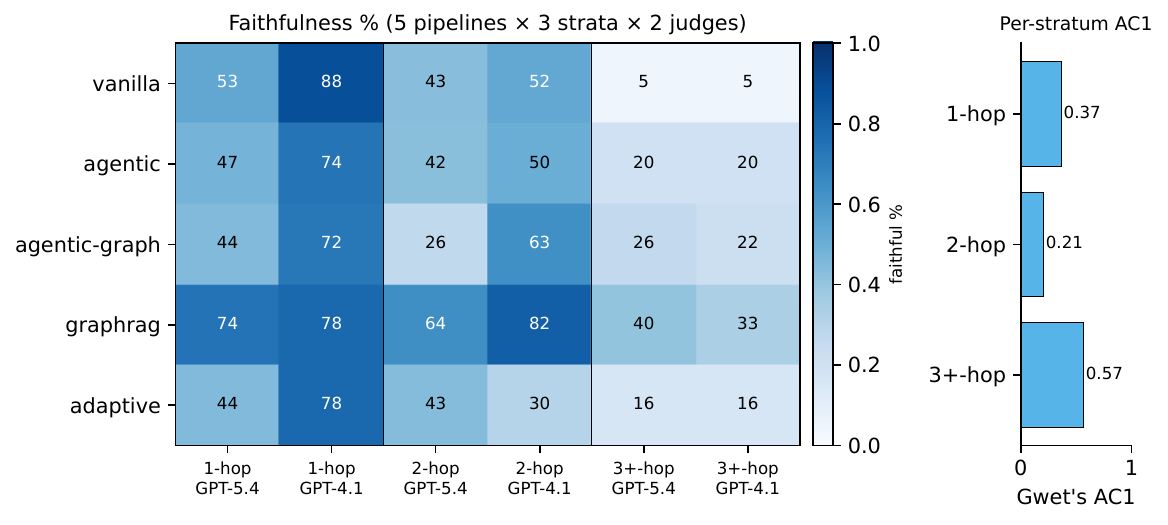}
  \caption{Multi-judge faithfulness on the v2 main matrix.
    \textit{Left}: faithfulness fraction (5 pipelines $\times$ 3 hop
    strata $\times$ 2 judges); the GraphRAG row shows the monotonic
    decline driving C2b on DO-178C. \textit{Right}: per-stratum Gwet
    AC1, showing the $\kappa$ paradox at 3+-hop (AC1${=}0.57$,
    $\kappa{<}0.10$; cf.\ Table~\ref{tab:judges}). v3 and MuSiQue:
    Table~\ref{tab:pathology}.}
  \Description{Heatmap of faithfulness percentages for five RAG
    pipelines (rows: vanilla, agentic, agentic-graph, graphrag,
    adaptive) across three hop strata times two judges (columns:
    1-hop GPT-5.4, 1-hop GPT-4.1, 2-hop GPT-5.4, 2-hop GPT-4.1,
    3+-hop GPT-5.4, 3+-hop GPT-4.1) on the v2 main matrix. The
    graphrag row darkens monotonically left-to-right under both
    judges. A side panel shows per-stratum Gwet AC1 bars.}
  \label{fig:faithheatmap}
\end{figure}

\subsection{C2: Context Flooding, Citation Selectivity, and Faithfulness}
\label{sec:c2}
\noindent\textbf{(C2a) The walk floods the context; the synthesizer
filters it} (Table~\ref{tab:pathology}, top). Across embedder \emph{and}
corpus swaps, GraphRAG's walk fills 11--15 context slots at context
precision 0.12--0.23 -- five in six retrieved chunks are off-gold. The
synthesizer then cites only 3.4--5.0 IDs per answer, at citation
precision 0.48--0.65: a 3--5$\times$ precision enrichment over its own
context. GraphRAG's $F_1$ edge is therefore a \emph{recall} effect
(citation recall 0.68--0.74 vs.\ vanilla's 0.46--0.65): the walk
surfaces gold chunks dense retrieval misses, and the synthesizer
declines to cite most of the noise that rides along. Evaluations that
score the retrieved set as the attribution set -- as some GraphRAG
comparisons do -- measure the flooding and miss the filtering, and
invert the ranking: by context precision GraphRAG is the worst pipeline
in every setting; by answer citations it is the best or tied-best.

\noindent\textbf{(C2b) The faithfulness consequence is corpus-conditional.}
The bottom block shows the split. On DO-178C, GraphRAG
faithfulness declines with hop distance: Cochran--Armitage trend
$p{=}0.040$ (GPT-5.4) and $p{=}0.008$ (GPT-4.1) under the local
embedder, $p{<}10^{-4}$ under Azure (GPT-4.1); only GPT-5.4-under-Azure
is flat ($p{=}0.51$). An
independent 300-tuple replication batch per embedder, drawn eleven
weeks later, reproduces the decline under GPT-5.4 in both embedders
($0.71{\to}0.20$, $p{=}0.002$ local; $0.79{\to}0.00$, $p{<}10^{-4}$
Azure) while GPT-4.1 -- by then judging 93--95\% of
everything faithful -- shows no trend ($p{=}0.89$ and $0.30$). The decline thus
replicates across batches, but \emph{which judge carries it swaps
with the judging date} (cf.\ C3). On MuSiQue neither judge collapses:
GPT-5.4 \emph{rises} (n.s., $p{=}0.068$) while GPT-4.1 declines mildly
from a 94\% base to 80\% ($p{=}0.017$), and both 3+-hop endpoints sit
far above the DO-178C endpoints (58\%/80\% vs.\ 40\%/33\%). The same
retrieval behavior thus has different faithfulness consequences on
different corpora; we hypothesize the mechanism in
\S\ref{sec:discussion}.

\begin{table}[t]
  \centering
  \caption{Judge fragility on the same 300 paired tuples. Top: inter-judge agreement halves under embedder swap (v2 local $\to$ v3 Azure). Bottom: same-judge controls --- test--retest (noise floor), embedder swap (41\% flip; GPT-4.1 more stable on that axis), and eleven-week re-judges of frozen inputs (temporal drift; GPT-4.1 at chance with itself).}
  \label{tab:judges}
  \small
  \setlength{\tabcolsep}{4pt}
  \resizebox{\columnwidth}{!}{%
  \begin{tabular}{l rrrr}
    \toprule
    \textit{Inter-judge $\kappa$} & 1-hop & 2-hop & 3+-hop & overall \\
    \midrule
    v2 main & 0.28 & 0.22 & 0.04 & 0.30 \\
    v3 main (pinned) & 0.27 & 0.07 & 0.07 & 0.17 \\
    MuSiQue (600 rows) & 0.00 & 0.01 & 0.07 & 0.02 \\
    \midrule
    \textit{Gwet's AC1} & 1-hop & 2-hop & 3+-hop & overall \\
    \midrule
    v2 main & 0.37 & 0.21 & 0.57 & 0.31 \\
    v3 main (pinned) & 0.22 & 0.07 & 0.35 & 0.17 \\
    \midrule
    \multicolumn{5}{l}{\textit{Same-judge controls (paired 300 tuples)}} \\
    \midrule
    GPT-5.4 test--retest (same day) & \multicolumn{4}{r}{$\kappa = 0.764$, raw agr.\ 0.88} \\
    GPT-5.4 across embedders (v2 vs v3) & \multicolumn{4}{r}{$\kappa = 0.137$, raw agr.\ 0.59} \\
    GPT-4.1 across embedders (v2 vs v3) & \multicolumn{4}{r}{$\kappa = 0.480$, raw agr.\ 0.74} \\
    GPT-5.4 same input, 11 wk apart & \multicolumn{4}{r}{$\kappa = 0.138$, raw agr.\ 0.56} \\
    GPT-4.1 same input, 11 wk apart & \multicolumn{4}{r}{$\kappa = -0.05\,/\,-0.00$ (v2/v3), raw agr.\ 0.48\,/\,0.51} \\
    \bottomrule
  \end{tabular}%
  }
\end{table}

\subsection{C3: Single-Judge Faithfulness is Retrieval-State-Fragile}
\label{sec:c3}
\looseness=-1
Table~\ref{tab:judges} reports inter-judge agreement on the paired
300-tuple subset and three same-judge controls. Inter-judge $\kappa$
between GPT-5.4 and GPT-4.1 halves under embedder swap on identical
tuples (0.30 $\to$ 0.17 overall); McNemar rejects symmetric disagreement
(exact $p{<}0.001$; GPT-4.1 systematically more lenient). On MuSiQue
the same pair agrees at chance level ($\kappa{=}0.02$, raw agreement
0.48, Gwet's AC1 0.08 -- genuine disagreement, not a prevalence
artifact; GPT-4.1 judges 88\% faithful vs.\ GPT-5.4's 47\%): C2b's
cross-corpus statements are per judge, each independently showing no
collapse. The same-judge controls carry the central result: same-input
same-day test--retest gives $\kappa{=}0.76$ (12\% of verdicts flip on
resampling alone), while re-judging the same tuples after an embedder
swap gives $\kappa{=}0.14$ (41\% flip). The swap changes retrieved contexts
and answers on 94\% of tuples: the gap thus measures verdict
sensitivity to \emph{upstream retrieval state} (output drift and judge
response jointly) against a known noise floor. A judge-specific
component remains: GPT-4.1 flips only 26\% ($\kappa{=}0.48$). A final
control: re-judging the identical 300 inputs eleven weeks later agrees
with the earlier verdicts at only $\kappa{=}0.14$ ($+15$\,pp leniency
shift). Sampling noise cannot explain this: under a stationary judge,
cross-date agreement should sit at the same-day test--retest level
(0.88); the observed 0.56 rejects stationarity (binomial
$p{<}10^{-44}$). The same control for GPT-4.1 is starker: re-judging
the identical tuples eleven weeks apart gives $\kappa{=}{-}0.05/{-}0.00$
(v2/v3) -- chance-level agreement with its own earlier verdicts, a
$+41$\,pp leniency shift. The judge that is more stable under
embedder swap is the less stable across time; stability on one axis
does not transfer to the other. The replication batches (C2b) show what that drift does at
scale: eleven weeks on, GPT-4.1 marks 93--95\% of fresh same-distribution
tuples faithful, inter-judge $\kappa$ falls to $0.05/0.01$
(local/Azure), and the trend GPT-4.1 itself had established in the
original batches disappears.

\begin{table}[t]
  \centering
  \caption{V2 router replication and Azure-embedder boost. Same 14-feature schema (11 hand text features, 3 PCs of the query embedding); only the embedder differs. Text features remain mostly inert (4/11 non-zero), so the +0.08 macro-$F_1$ gain is attributable to dense-embedding hop-decodability.}
  \label{tab:router}
  \small
  \resizebox{\columnwidth}{!}{%
  \begin{tabular}{l rr}
    \toprule
     & v2 (local 384d $\to$ PCA-3) & v3 (Azure 1536d $\to$ PCA-3) \\
    \midrule
    50-fold macro-$F_1$ & $0.78 \pm 0.08$ & $\mathbf{0.86 \pm 0.05}$ \\
    Adaptive-V2 overall $F_1$ & 0.559 & 0.569 \\
    Adaptive-V1 overall $F_1$ & 0.439 & 0.468 \\
    Oracle overall $F_1$ & 0.641 & 0.652 \\
    V2 $-$ V1 (absolute) & $+0.120$ & $+0.101$ \\
    Gap closure (V1$\to$Oracle) & 59.4\% & 54.8\% \\
    Holm-Wilcoxon strata sig & 3/3 & 2/3 \\
    \bottomrule
  \end{tabular}}
\end{table}

\paragraph{C4 (supporting): Dense embeddings alone classify hop.}
Table~\ref{tab:router} shows the V2 router replication. The 50-fold
macro-$F_1$ improves $0.78\pm0.08\to 0.86\pm0.05$ under embedder swap,
and Adaptive-V2 closes 59\%/55\% of the V1$\to$Oracle gap
(local/Azure), with Holm-significant per-stratum gains in 3/3 strata
locally and 2/3 under Azure (the 3+-hop routing targets are
statistically tied there). Only 4 of the 11 hand text
features are ever non-zero on the 296 queries; the rest are inert. The router therefore operates almost
exclusively on the 3 query-embedding PCs, suggesting hop distance is
largely an embedding-decodable property of the query text in our
generated query set; transfer to human-authored queries is untested.

\section{Discussion}
\label{sec:discussion}

\paragraph{Why does flooding's harm depend on the corpus?}
GraphRAG's traversal expands candidates well beyond what dense
retrieval surfaces, and the synthesizer filters most of the noise out
of its citations (C2a). What it cannot filter is the influence of that
noise on the answer itself: faithfulness is judged against the full
retrieved block, and whether flooding hurts depends on what the edges
connect. \textit{derives\_from} and \textit{references} expansion on
DO-178C drags in cross-module artifacts whose claims contradict the
question, while Wikipedia expansion stays topically adjacent and
typically satisfies the judge's support criterion. We expect the same
coupling wherever typed-edge expansion crosses requirements that
contradict one another.

\paragraph{Implications for LLM-judge practice.}
Same-judge self-$\kappa$ across embedders is $0.137$, against a
same-day test--retest floor of $0.76$; on frozen inputs eleven weeks
apart, self-$\kappa$ is $0.14$ (GPT-5.4) and $-0.05$ (GPT-4.1). A single-judge comparison across
retrieval modules therefore reports embedder- and date-conditional
verdicts, not invariant faithfulness. Multi-judge ensembles mitigate
this only partially, since inter-judge $\kappa$ itself halves under
embedder swap. Paired-tuple judging under a fixed embedder is the
least that is needed for comparable faithfulness numbers.

\paragraph{Threats to validity.}
\looseness=-1
\textit{Construct.} ALCE-style $F_1$ does not measure rationale quality;
we pair it with multi-judge faithfulness, while acknowledging LLM-judge
bias~\cite{selfpref_bias2024,wallat2024}; C3 quantifies this directly.
Both judges share a vendor, and cross-vendor replication is future work.
\textit{External.} The DO-178C-style corpus is a single synthetic
aerospace dataset, generated by the same model family that answers and
judges; absolute $F_1$ levels characterize this single
regulated-domain setting and may not transfer. A generator-swap
control (GPT-4.1 re-synthesizing 332 rows on frozen retrievals)
reproduces the architecture ordering with slightly higher $F_1$, and
its answers show the same hop-wise faithfulness decline under both
judges, so neither citation nor faithfulness results are
GPT-5.4-specific; the corpus-author leg of the circularity remains. The MuSiQue
cross-corpus replication (C2a, C2b) is the main
external-validity check for the universality claim; cross-corpus
replication of C1 is left for the journal extension. Our graph pipeline
is one typed-edge walk implementation; other GraphRAG-family systems
may flood or filter differently.
\textit{Statistical.} Per-stratum $n\!\in\![95,111]$ is at the lower
edge of BCa stability under heavy skew; a percentile fallback shows no
qualitative reversal. \textit{Reranker / router engineering.} No cross-encoder
reranker is included (orthogonal axis); C4 is a diagnostic instrument,
not a production router.

\section{Conclusion}
\label{sec:conclusion}
Across a triple-robustness (embedder $\times$ corpus $\times$ judge)
analysis of five RAG architectures: \emph{graph expansion floods the
context but the synthesizer cites selectively, and moving the
measurement point from context to answer inverts the architecture
ranking}; \emph{the faithfulness consequence of flooding is
corpus-conditional}; \emph{stratum winners are embedder-robust but
corpus-conditional}; and \emph{single-judge verdicts are
retrieval-state- and date-fragile}. Triple-robustness -- and an
explicit choice of citation-measurement point -- is the minimum bar
for RAG architecture claims.

\bibliographystyle{ACM-Reference-Format}
\bibliography{refs}

\end{document}